\documentclass[runningheads]{llncs}
\usepackage{graphicx}
\usepackage{todonotes}
\usepackage{enumitem,amssymb}
\usepackage{todonotes}
\usepackage{pifont}

\usepackage{url}
\usepackage{color, colortbl}
\usepackage{booktabs}
\usepackage{fixltx2e}
\usepackage{floatrow}  
\usepackage{blindtext}  
\usepackage{array, caption, floatrow, makecell, booktabs}  
\usepackage{multirow}  
\usepackage{rotating}  
\usepackage{color,soul}

\usepackage{fdsymbol}

\usepackage{amsmath}
\usepackage{dsfont}
\usepackage{mathtools}

\usepackage{algorithmic}

\usepackage{amsmath}
\usepackage{xparse}

\ExplSyntaxOn
\NewDocumentCommand{\ucgreek}{m}
 {\str_case:nn { #1 } {
    {A}{\mathrm{A}} {B}{\mathrm{B}} {C}{\Sigma} {D}{\Delta} {E}{\mathrm{E}} 
    {F}{\Phi} {G}{\Gamma} {H}{\mathrm{H}} {I}{\mathrm{I}} {J}{\Theta} {K}{\mathrm{K}} 
    {L}{\Lambda} {M}{\mathrm{M}} {N}{\mathrm{N}} {O}{\mathrm{O}} {P}{\Pi}
    {Q}{\mathrm{X}} {R}{\mathrm{P}} {S}{\Sigma} {T}{\mathrm{T}} {U}{\Upsilon} 
    {W}{\Omega} {X}{\Xi} {Y}{\Psi} {Z}{\mathrm{Z}}
}}
\NewDocumentCommand{\lcgreek}{m}
 {\str_case:nn { #1 }
   {{a}{\alpha} {b}{\beta} {c}{\varsigma} {d}{\delta} 
    {e}{\varepsilon} {f}{\varphi} {g}{\gamma} {h}{\eta} {i}{\iota}
    {j}{\vartheta} {k}{\kappa} {l}{\lambda} {m}{\mu} {n}{\nu} {o}{o}
    {p}{\pi} {q}{\chi} {r}{\rho} {s}{\sigma} {t}{\tau} {u}{\upsilon} 
    {w}{\omega} {x}{\xi} {y}{\psi} {z}{\zeta}
}}
\ExplSyntaxOff

\newcommand{\mathimage}[1]{\mathbf{#1}}

\newcommand{\mathscal}[1]{\lowercase{\textit{#1}}}
\newcommand{\mathtensor}[1]{\mathrm{\uppercase{#1}}}
\newcommand{\mathdistrib}[1]{\mathcal{#1}}
\newcommand{\mathfunc}[1]{\uppercase{\ucgreek{#1}}} 

\usepackage{pifont}

\usepackage[normalem]{ulem}

\newcommand{\darkbluetext}[1]{\color[RGB]{0,10,133}{#1}\color{black}}
\newcommand{\browntext}[1]{\color[RGB]{117,20,12}{#1}\color{black}}

\definecolor{lightgray}{rgb}{0.92,0.92,0.92}
\newcolumntype{g}{>{\columncolor{lightgray}}c}
\newcolumntype{?}{!{\vrule width 1.5pt}}  

\newfloatcommand{capbtabbox}{table}[][\FBwidth]

\newlist{todolist}{itemize}{2}
\setlist[todolist]{label=$\square$}

\graphicspath{{figures/}}
\newcommand{\projectpage}{\url{https://vios-s.github.io/adversarial-test-time-training}}

\begin{document}


\title{Stop Throwing Away Discriminators! \\Re-using Adversaries for Test-Time Training}
\titlerunning{Stop Throwing Away Discriminators! Re-using Adversaries for TTT} 
\author{Gabriele Valvano\inst{1, 2} 
        Andrea Leo\inst{1} \and
        Sotirios A. Tsaftaris\inst{2}}
        
\authorrunning{G. Valvano et al.} 
\institute{
    IMT School for Advanced Studies Lucca, Lucca 55100 LU, Italy 
    \and School of Engineering, University of Edinburgh, Edinburgh EH9 3FB, UK
} 

\maketitle              

\begin{abstract}
Thanks to their ability to learn data distributions without requiring paired data, Generative Adversarial Networks (GANs) have become an integral part of many computer vision methods, including those developed for medical image segmentation. 
These methods jointly train a segmentor and an adversarial mask discriminator, which provides a data-driven shape prior. 
At inference, the discriminator is discarded, and only the segmentor is used to predict label maps on test images. But should we discard the discriminator?
Here, we argue that the life cycle of adversarial discriminators should not end after training. On the contrary, training stable GANs produces powerful shape priors that we can use to \textit{correct} segmentor mistakes at inference.
To achieve this, we develop stable mask discriminators that do not overfit or catastrophically forget. At test time, we fine-tune the segmentor on each individual test instance until it satisfies the learned shape prior. 
Our method is simple to implement and increases model performance. Moreover, it opens new directions for re-using mask discriminators at inference.
We release the code used for the experiments at~\projectpage.

\end{abstract}

\section{Introduction}
Semi- and weakly-supervised learning are emerging paradigms for image segmentation \cite{review2019notsosup,review2020embracing}, often involving adversarial training \cite{goodfellow2014generative} when annotations are sparse or missing.
Adversarial training involves two simultaneously trained networks: one focusing on an image generation task, and the other learning to tell apart generated images from real ones. In semantic segmentation, it is standard practice to condition the generator, also termed segmentor, on an input image and optimise it to output realistic and accurate segmentation masks. After training, the discriminator is discarded and the segmentor used for inference. 

Unfortunately, segmentors may under-perform and make errors whenever the test data fall outside the training data distribution (e.g., because acquired with a different scanner or belonging to a different population study). Here we propose a simple mechanism to detect and correct such errors in an end-to-end fashion, re-using components already developed during training. 

We embrace an emerging paradigm \cite{sun2020test,wang2020fully,karani2021test,he2021autoencoder} where a model is fine-tuned on individual test instances without requiring access to other data nor labels. We propose strategies that permit \textit{recycling} an adversarial mask discriminator during inference, thus introducing a data-driven shape prior to correct predictions.  
Motivated by recent findings of Asano et al. \cite{asano2019critical}, reporting that we can effectively train the early layers' weights of a CNN with just one image, we propose to tune them on a per-testing instance to minimise an adversarial loss. Lastly, contrary to standard post-processing operations, our method can potentially learn from a continuous stream of data \cite{sun2020test}. 
Our \textbf{contributions} are:
\textbf{1)} to the best of our knowledge, this is the first attempt to use adversarial mask discriminators to \textit{detect} and \textit{correct} segmentation mistakes during inference; \textbf{2)} we define specific assumptions (and show how to satisfy them) to make the discriminators useful after training; \textbf{3)} we report performance increase on several medical datasets.

\section{Related Work}
\noindent\textbf{Learning from Test Samples.}
In our work, we use a discriminator to tune a segmentor on the individual \textit{test} images until it predicts realistic masks. The idea of fine-tuning a model on the test samples has recently been introduced by Sun et al. \cite{sun2020test} with the name of Test-time Training (TTT). TTT optimises a model by jointly minimising a supervised and an auxiliary self-supervised loss on a training set, such as detecting the rotation angle of an input image. 
At inference, TTT fine-tunes the model to minimise the auxiliary loss on the individual test instances, thus adapting to potential distribution shifts.
Although the model was successful for classification, the authors admit that designing a well-suited auxiliary task is non-trivial. 
For example, predicting a rotation angle may be less effective for medical image segmentation, where images have different acquisition geometries. 
Moreover, Sun et al. only test their model ``simulating" domain shifts with hand-crafted image corruptions (e.g., noise and blurring) without investigating if TTT can improve segmentation performance.

Following this seminal work, Wang et al. \cite{wang2020fully} suggested tuning an adaptor network to minimise the test prediction entropy. %
Unfortunately, CNNs usually make low-entropy overly-confident predictions \cite{guo2017calibration}, and entropy minimisation could be sub-optimal for segmentation. More crucially, Wang et al. rely on having access to the \textit{entire} test-set to do the fine-tuning.

Karani et al. \cite{karani2021test} recently proposed Test-time Adaptable Neural Networks to extend TTT for image segmentation using a pre-trained mask denoising autoencoder (DAE). At inference, they compute a reconstruction error between the mask generated by a segmentor and its auto-encoded version predicted by the DAE. 
This error constitutes a test-time loss used to fine-tune a small adaptor CNN in front of the segmentor. 
Once tuned, the adaptor maps the individual test images onto a normalised space which overcomes domain shifts problems for the segmentor. 
A limitation of this approach is the need to train the mask DAE separately. On the contrary, GANs learn the shape prior and optimise the segmentor in an \textit{end-to-end} fashion. 
Moreover, tuning the model with a convolutional encoder (the discriminator) rather than an autoencoder has advantages in terms of occupied memory and is faster at inference.
Herein, we show that improving performance using a discriminator is also possible and, at the same time, we open a new research direction toward learning re-usable discriminators.

\noindent\textbf{Shape Priors in Deep Learning for Medical Image Segmentation.} 
Incorporating prior knowledge about organ shapes is not uncommon in medical imaging \cite{nosrati2016incorporating}. 
Several methods introduced shape priors to regularise the training of a segmentor using penalties~\cite{kervadec2019constrained,clough2019topological}, autoencoders \cite{oktay2017anatomically,dalca2018anatomical}, atlases \cite{dalca2019unsupervised}, and adversarial learning \cite{yi2019generative,valvano2021learning}.
Others included shape priors for post-processing, fixing prediction mistakes \cite{painchaud2019cardiac,larrazabal2020post}. 
GANs have become a popular way of introducing shape priors for image segmentation \cite{yi2019generative}, with the advantage of: i) learning the prior directly from data; ii) having a simple model that works well for semi- and weakly-supervised learning; and iii) learning the prior while also training the segmentor, instead of in two separate steps (as happens for autoencoders).

\noindent\textbf{Re-using Adversarial Discriminators.}
Re-using pre-trained discriminators was proposed to obtain features extractors for transfer learning \cite{radford2015unsupervised,donahue2016adversarial,mao2019discriminator}, or anomaly detectors \cite{zenati2018adversarially,ngo2019fence}. 
To the best of our knowledge, their (re-)use to detect segmentor mistakes during inference remains unexplored. We are also not aware of previous use of discriminators for test-time tuning of a segmentor.

\begin{figure}[t]
    \centering
    \includegraphics[width=0.9\linewidth]{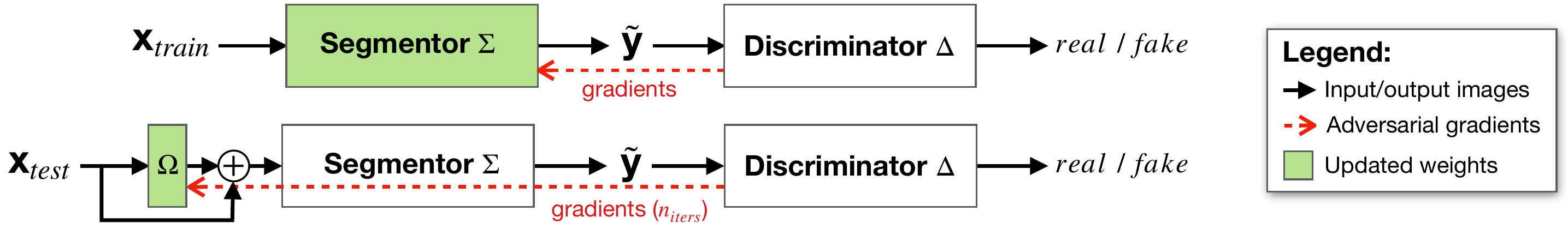}
    \caption{
    We re-use GAN discriminators to correct segmentation predictions at inference. The key to our success is training stable and re-usable discriminators, as we detail in Section~\ref{subsec:challenges_and_solutions}. At inference, we tune a small convolutional block $\mathfunc{w}$ on each test image $\mathimage{x}$, independently, until the predicted mask $\mathimage{\tilde{y}}$ satisfies the adversarial shape prior. We only need a single test sample to do the fine-tuning.%
    }
    \label{fig:method_overview}
\end{figure}
\section{Method}

As summarised in Fig.~\ref{fig:method_overview}, we consider two stages: i) standard adversarial training; and ii) at inference, image-specific tuning of a small adaptor CNN ($\mathfunc{w}$) in front of the trained segmentor. 
In the first stage, we optimise a segmentor $\mathfunc{s}$ to minimise a supervised cost on the annotated data and an adversarial cost on a set of unpaired images. Meanwhile, we train the discriminator $\mathfunc{d}$ to distinguish real from predicted masks. 
At inference, for each test-image, we tune the adaptor $\mathfunc{w}$ using the (unsupervised) adversarial loss, and improve performance. For $\mathfunc{s}$ and $\mathfunc{w}$ we use architectures that proved to be effective in segmentation tasks \cite{ronneberger2015u,karani2021test}, while we leave exploring alternative architectures as future work.

Obtaining discriminators re-usable at inference is not trivial and requires specific solutions to overcome crucial challenges. These solutions, with our optimisation strategy and model design, are one major contribution of this work. 

In the following, we will use italic lowercase letters to denote scalars $\mathscal{s}$, and bold lowercase for 2D images $\mathimage{x} \in \mathds{R}^{n \times m}$, where $n, m \in \mathbb{N}$ are the height and width of the image, respectively. 
Lastly, we adopt capital Greek letters for functions $\mathfunc{f}$.

\subsection{Re-usable Discriminators: Challenges and Proposed Solutions}\label{subsec:challenges_and_solutions}
\noindent\textbf{Challenge 1.} To obtain a re-usable discriminator $\mathfunc{d}$, we must prevent it from \textit{overfitting} and \textit{catastrophically forget}, or its predictions on the masks generated during inference will not be reliable. Generally speaking, this is a challenging task because: GANs can easily memorise data if trained for too long \cite{nagarajan2018theoretical}.\footnote{Memorisation can also happen just in the discriminator. In fact, contrarily to the segmentors, we do not use any additional supervised cost to regularise the discriminator training. We show how to detect memorisation from the losses in the Supplemental.
} Moreover, the discriminator may forget how unrealistic segmentation masks look like after the segmentor training has converged \cite{shrivastava2017learning}.
Although $\mathfunc{d}$ may work well at training in these cases, it would not generalise to the test data, as we explain below.

If properly trained, a segmentor $\mathfunc{s}$ predicts \textit{realistic} segmentation masks in the latest stages of training. Thus, in standard GANs, we stop training while optimising $\mathfunc{d}$ to tell apart \textit{real} from more and more \textit{real-looking} masks. 
At convergence, this becomes similar to training the discriminator using only \textit{real} images and labelling them as \textit{real} half the times, as \textit{fake} the other half. 
At this point, gradients become uninformative, and the discriminator collapses to one of the following cases: 
\textbf{i)} it always predicts its equilibrium point (which in vanilla GANs is the number 0.5, equidistant from the labels \textit{real}: 1, \textit{fake}: 0) but it can still detect unrealistic images; 
\textbf{ii)} it predicts the equilibrium point independently of the input image, forgetting what \textit{fake} samples look like \cite{shrivastava2017learning,kim2018memorization}; 
or \textbf{iii)} it memorises the real masks (which, differently from the generated ones, appear unchanged since the beginning of training) and it always classifies them as \textit{real}, while classifying \textit{any other input} as \textit{fake}. 
It is crucial to prevent the behaviours \textbf{ii)} and \textbf{iii)} to have a re-usable discriminator. For this reason, we use:
\begin{itemize}
    \item \textit{Fake anchors}: we ensure to expose the discriminator to unrealistic masks (labelled as \textit{fake}) until the end of training. In particular, we train $\mathfunc{d}$ using real masks $\mathimage{y}$, predicted masks $\tilde{\mathimage{y}}$, and corrupted masks $\mathimage{y}_{corr}$. We obtain $\mathimage{y}_{corr}$ by randomly swapping squared patches within the image\footnote{We use patches having size equal to 10\% of the image size.} and adding binary noise to the real masks, as this proved to be a fast and effective strategy to learn robust shape priors in autoencoders \cite{karani2021test}.
    While, towards the end of the training, the discriminator may not distinguish $\mathimage{y}$ from the real-looking $\tilde{\mathimage{y}}$, the exposure to $\mathimage{y}_{corr}$ will prevent forgetting how unrealistic masks look like, providing informative gradients until we stop training.
\end{itemize}

\noindent\textbf{Challenge 2.} An additional challenge is to train \textit{stable} discriminators, which do not change much in the latest training epochs. In other words, we want small oscillations in the discriminator loss. This is necessary because we typically stop training using early stopping criteria on the segmentor loss. Therefore, we want to promote the optimisation of Lipschitz smooth discriminators, avoiding suddenly big gradient updates (thus leaving $\mathfunc{d}$ mostly unchanged between the last few training epochs).
To this end, we suggest using:
\begin{itemize}
    \item \textit{Smoothness Constraints}: we increase discriminator smoothness \cite{chu2020smoothness} through Spectral normalisation \cite{miyato2018spectral} and $tanh$ activations.
    \item \textit{Discriminator data augmentation}: consisting of random roto-translations, and Instance Noise \cite{sonderby2016amortised,muller2019does}, to map similar inputs to the same prediction label. We translate images up to 10\% of image pixels on both vertical and horizontal axes, and we rotate them between $0 \div \pi/2$. We generate noise using a Normal distribution with zero mean and 0.1 standard deviation.
\end{itemize}

\subsection{Architectures and Training Objectives for $\mathfunc{s}$ and $\mathfunc{d}$}
We use a UNet \cite{ronneberger2015u} segmentor with batch normalisation \cite{ioffe2015batch}. Given an input image $\mathimage{x}$, the segmentor $\mathfunc{s}$ predicts a multi-channel label map $\tilde{\mathimage{y}}=\mathfunc{s}(\mathimage{x})$. For the annotated images, we minimise the supervised weighted cross-entropy loss: 
\begin{align}\label{eq:sup_loss}
        \mathcal{L}(\mathfunc{s})  
                = - \sum\nolimits_{\mathscal{i}=1}^{\mathscal{c}} 
                \mathscal{w}_i \cdot \mathimage{y}_i \log(\tilde{\mathimage{y}}_i)
\end{align}
where $i$ is a class index, $\mathscal{c}$ the number of classes, and $\mathscal{w}_i$ a class scaling factor used to address the class imbalance problem. The value $\mathscal{w}_i = 1 - \mathscal{n}_i/\mathscal{n}_{tot}$ considers both the total number of pixels $\mathscal{n}_{tot}$ and the number of pixels $n_i$ with label $\mathscal{i}$.

As discriminator $\mathfunc{d}$, we use a convolutional encoder, processing the predicted masks with a series of 5 convolutional layers. Layers use a number of $4 \times 4$ filters following the series: 32, 64, 128, 256, 512. After the first two layers, we downsample the features maps using a stride of 2.
We increase smoothness according to Section~\ref{subsec:challenges_and_solutions}.
Finally, a fully-connected layer integrates the extracted features and predicts a scalar linear output, used to compute the adversarial losses \cite{mao2018effectiveness}:
\begin{equation}\label{eq:adv_loss}
\begin{split}
         & \min_{\mathfunc{d}} \bigg\{
         \mathcal{V}_{LS}(\mathfunc{d}) = 
            \frac{1}{2} E_{\mathimage{y} \sim \mathdistrib{Y}}[(\mathfunc{d}(\mathimage{y}) - 1)^2] 
            + 
            \frac{1}{2} E_{\mathimage{x} \sim \mathdistrib{X}}[(\mathfunc{d}(\mathfunc{s}(\mathimage{x})) + 1)^2] \bigg\}
        \\ &
        \min_{\mathfunc{s}} \bigg\{
        \mathcal{V}_{LS}(\mathfunc{s}) = 
            \frac{1}{2} E_{\mathimage{x} \sim \mathdistrib{X}}[(\mathfunc{d}(\mathfunc{s}(\mathimage{x})))^2] \bigg\}
\end{split}
\end{equation}
where $-1$ and $+1$ are the labels for \textit{fake} and \textit{real} images, respectively.
As in standard adversarial semi-supervised training, we alternately minimise eq.~\ref{eq:sup_loss} on a batch of annotated images and eq.~\ref{eq:adv_loss} on a batch of unpaired images and unpaired masks. 
We use Adam optimiser \cite{kingma2014adam}, learning rate: $10^{-4}$, and batch size: 12. Tra\-ining proceeds until the segmentation loss stops decreasing on a validation set.

\subsection{Adversarial Test-Time Training: Adapting $\mathfunc{w}$}
At inference, we do not fine-tune the whole segmentor but only adapt a few convolutional layers at its input. These layers are, according to \cite{asano2019critical}, the most suited for one-shot learning. By keeping the deeper layers of $\mathfunc{s}$ unchanged, we also limit the segmentor flexibility and let it adapt only to changes at lower abstraction levels, ultimately preventing trivial solutions.
Thus, we include a shallow convolutional residual block (adaptor $\mathfunc{w}$) in front of the segmentor, that we tune on the individual test images by minimising $\mathcal{V}_{LS}(\mathfunc{s} \equiv \mathfunc{w})$ for $\mathscal{n}_{iter}$ iterations.
The adaptor is the same as in \cite{karani2021test} and has 3 convolutional layers with 16 $3\times3$ kernels and activation $e^{-\mathtensor{t}^2 / \mathscal{s}^2}$, being $\mathtensor{t}$ an input tensor and $\mathscal{s}$ a train\-a\-ble scaling parameter, initialised as 0 and optimised at inference. 
After tuning $\mathfunc{w}$, the input to the segmentor is an augmented version of $\mathimage{x}$ which can be more easily classified. We show qualitative examples in Fig.~\ref{fig:adaptation_and_segmentation} and in the Supplemental.

\section{Experiments}

\begin{figure}[t]
    \centering
    \includegraphics[width=0.65\linewidth]{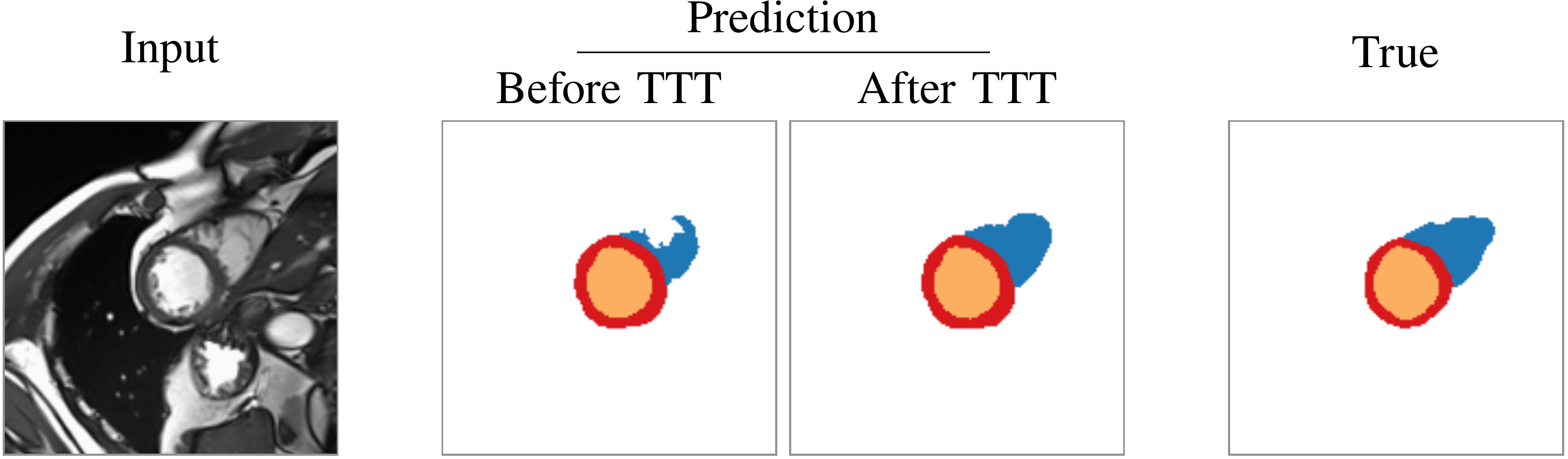}
    \caption{Effect of Test-time Training (TTT). Re-using a discriminator at inference, we optimise a small input adaptor $\mathfunc{w}$ until the predicted mask becomes realistic. We report additional examples in the Supplemental.}
    \label{fig:adaptation_and_segmentation}
\end{figure}

\begin{figure}[t]
    \centering
    \includegraphics[width=0.75\linewidth]{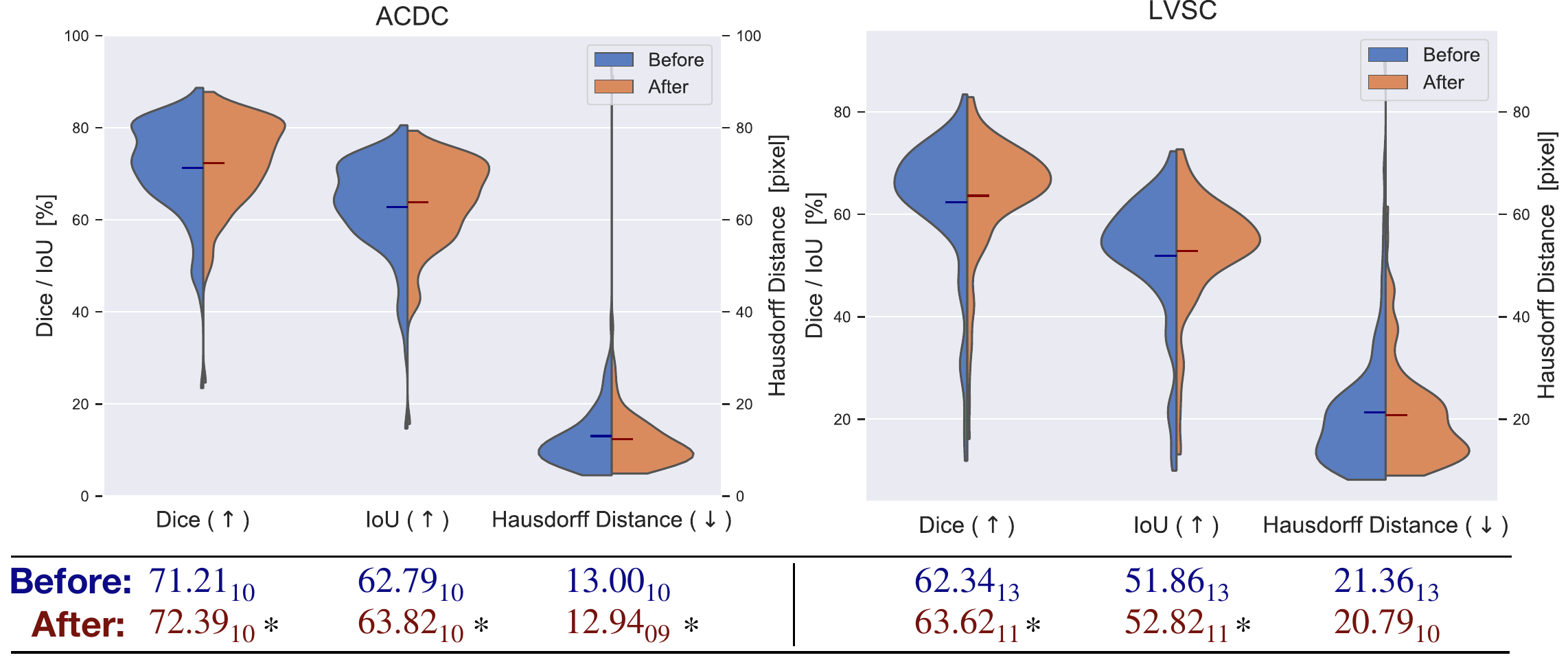}
    \caption{Dice (\textuparrow), IoU (\textuparrow) and Hausdorff distance (\textdownarrow) obtained \darkbluetext{\textbf{before} } and \browntext{\textbf{after} } tuning the segmentor on the individual test instances. Arrows show metric improvement directions.
    Under each violin plot, we also report the average performance (standard deviation as subscript). 
    We always improve the metrics, also in the worst-case scenarios (bottom of the distribution tails for Dice and IoU, upper tails for Hausdorff distance). 
    Asterisks show statistical significance.
    }
    \label{fig:violin_plot}
\end{figure}

\paragraph{Data}
We consider two cardiac MRI datasets, described below.\\
\noindent \textbf{ACDC}~\cite{bernard2018deep} has multi-scanner images from 100 patients with manual annotations for right and left ventricle, and for left myocardium. We resample data to the average resolution: 1.51$mm^2$, and crop/pad them to $224\times224$ pixels. We standardise data using the patient-specific median and interquartile range. \\
\noindent \textbf{LVSC}~\cite{suinesiaputra2014collaborative} contains cardiac MRIs of 100 subjects, obtained with different scanners and imaging parameters. There are manual annotations for the left myocardium. We resample images to the average resolution of 1.45$mm^2$, and then crop or pad them to $224\times224$ pixel size. We normalise images as in ACDC.

\paragraph{Setup and Evaluation.} We divide datasets by patients, using groups of 40\% for training, 20\% for validation, and 40\% for the test set, respectively. Out of the 40\% training patients, we consider annotations for one fourth of the training subjects in ACDC and LVSC (10 patients). We treat the remaining data as unpaired and use them for adversarial training (eq.~\ref{eq:adv_loss}). 
Notice that the small training sets cannot fully represent the entire data distribution, leading to segmentation errors at inference. We will investigate dealing with larger distribution shifts (e.g. different scanners, etc.) in the future. 
We analyse performance increases obtained from adversarial Test-time Training. Inspired by \cite{karani2021test}, we also compare to using a DAE to drive the adaptation (DAEs learn the shape prior separately, we do it while training $\mathfunc{s}$). 
We do 3-fold cross-validation and measure performance comparing the predicted segmentation masks and the ground truth labels contained in the test set. We use Dice and IoU scores, and the Hausdorff distance. We assess statistical significance with the non-parametric Wilcoxon test ($p = 0.01$).

\begin{figure}[t]
    \centering
    \includegraphics[width=0.8\linewidth]{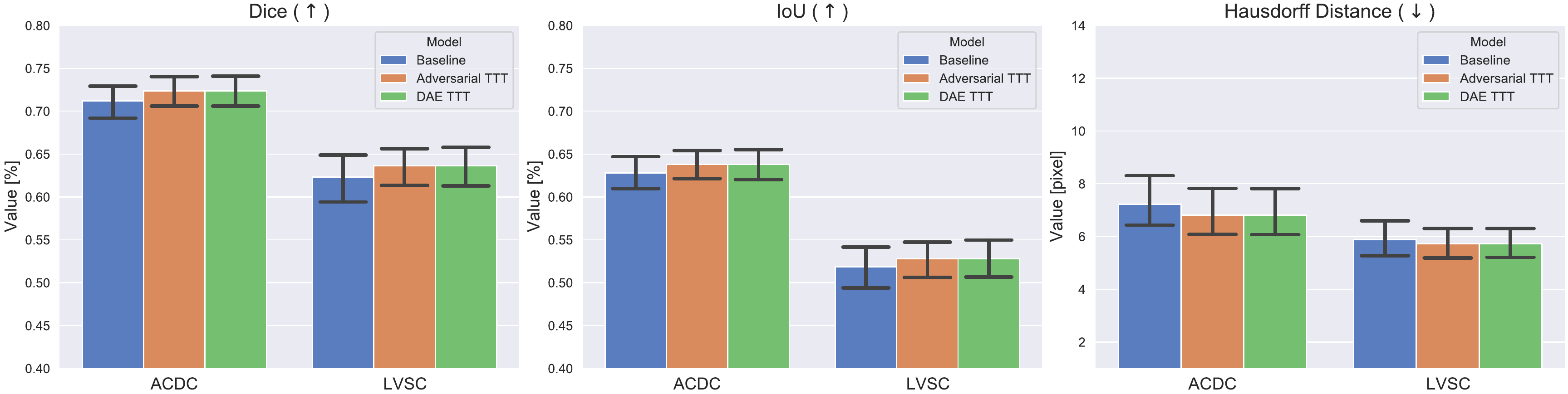}
    \caption{
    Test-time Training using an adversarial shape prior vs a prior learned by a pre-trained DAE. In the plots, ``Baseline" refers to standard inference of a GAN (i.e. without TTT). 
    Bar plots show average and 95\% confidence interval. 
    Both methods lead to similar improvements ($p=0.01$).
    }
    \label{fig:bar_plot}
\end{figure}

\subsection{Results and Discussion}
We show a qualitative example of test-time adaptation in Fig.~\ref{fig:adaptation_and_segmentation}. 
In Fig.~\ref{fig:violin_plot}, we represent segmentation performance with violin plots before and after Test-time Training. These plots show the whole distribution of performance values for images in the test set. We observe performance improvements on all metrics and datasets. Importantly, worst-case scenarios (bottom tails of violin plots, for Dice and IoU; top tails for Hausdorff distance) considerably improve,
reflecting the desired tendency to correct unrealistic segmentation masks that do not satisfy the learned adversarial shape prior. Qualitatively, we observed that the model removes scattered false positives and closes holes in the segmentation masks (see Fig. 1 in the Supplementary material). 

In Fig.~\ref{fig:bar_plot}, we compare the performance of our method vs using a shape prior separately learned by a DAE, inspired by \cite{karani2021test}. 
Our method achieves similar performance gains to a DAE (no statistically significant differences found), but it has the advantage of \textit{not} requiring a separate pre-training step. 

%

\begin{table}[t]
    \caption{
     Ablation Study. We compare the performance of a UNet; 
     a standard GAN; the GAN after adding: smoothness constraints (\#1), the proposed regularisation technique: \textit{fake anchors} (\#2), and Test-Time Training (\#3). Results are average (standard deviation as subscript) Dice scores on the ACDC test set.
    }\label{tab:ablation}
    \begin{tabular}{c|c|c|c|c}
    		
            UNet~ &
            ~ GAN ~ & 
            ~ GAN + \#1 ~ & 
    		~ GAN + \#1 + \#2 ~ & 
            ~ GAN + \#1 + \#2 + \#3 \\
    		\midrule
    		
    		70.1\textsubscript{13}  &
    		70.0\textsubscript{12}  &
    		70.9\textsubscript{11}  &
    		71.2\textsubscript{10}  &
    		72.4\textsubscript{10} \\
    		\bottomrule
    	\end{tabular}
\end{table}

Lastly, we perform an ablation study to analyse the effect regularising the model with \textit{smoothness constraints} and \textit{fake anchors}. As illustrated in Table~\ref{tab:ablation}, the techniques 
improves 
training and makes the adversarial shape prior stronger. As a result: i) the adversarial training leads to a better segmentor; and ii) the re-usable discriminator further increases model performance.

\paragraph{Computational Aspects}
The memory required to store the weights of the model is 90 MB. 
At inference, our method needs $\mathscal{n}_{iter}$ forward and backward passes to correct a segmentation.
This is slower than standard inference, where each image requires one forward pass. We find that $\mathscal{n}_{iter} > 0$ improves segmentation, but high values (e.g. $>100$) overfit the segmentor leading to worse performance. As a compromise, we use $\mathscal{n}_{iter}=50$ (with small temporal overhead: $\sim$10s/patient on a TITAN Xp GPU). %
This fixed iteration strategy is also used by previous work \cite{sun2020test,karani2021test,he2021autoencoder}, but using image-specific optimal $\mathscal{n}_{iter}$ would be useful and potentially increase performance. We leave automated strategies to set $\mathscal{n}_{iter}$ as future work.

\paragraph{Limitations}
We find that $\mathfunc{d}$ does not penalise wrong predictions that appear realistic but do not correspond to the input image. In fact, the discriminator only evaluates the predicted mask without considering the segmentor input. We highlight that this is also a limitation of \cite{karani2021test} and of all methods learning the shape prior only using unpaired masks. We expect that including also the image-related information would improve Test-time Training.


\section{Conclusion}
We demonstrated that by satisfying simple assumptions, it is possible to re-use adversarial discriminators during inference. In particular, we re-used a mask discriminator to detect and then correct segmentation mistakes made by a segmentor. The proposed method is simple and can be potentially applied to any GAN, increasing its test-time performance on the most challenging images. 

More broadly, the possibility of re-using adversarial discriminators to correct generator errors may open opportunities even outside image segmentation.
Given their flexibility and the ability to learn data-driven losses, GANs have been widely adopted in  medical imaging, from domain adaptation to image synthesis tasks \cite{yi2019generative}. With improved architectures and regularisation techniques \cite{kurach2019large,chu2020smoothness}, we believe adversarial networks will be even more popular in the future. 
In this context, training stable and re-usable discriminators opens opportunities for an all-round use of the GAN components.

\subsubsection{Acknowledgments}
This work was partially supported by the Alan Turing Institute (EPSRC grant EP/N510129/1). S.A. Tsaftaris acknowledges the support of Canon Medical and the Royal Academy of Engineering and the Research Chairs and Senior Research Fellowships scheme (grant RCSRF1819\textbackslash8\textbackslash25).

\bibliographystyle{splncs04}
\bibliography{references}
\end{document}


\title{Stop Throwing Away Discriminators! \\Re-using Adversaries for Test-Time Training \\ (Supplementary Material)}
\titlerunning{Supplementary Material} 

\author{}
\institute{}
\maketitle

\section{Additional Test Examples}
Below, we report examples of mistaken predictions and their corrections obtained after adversarial Test-time Training. We group pairs of examples by dataset.

\begin{figure}[h!]
    \centering
    \includegraphics[width=0.85\linewidth]{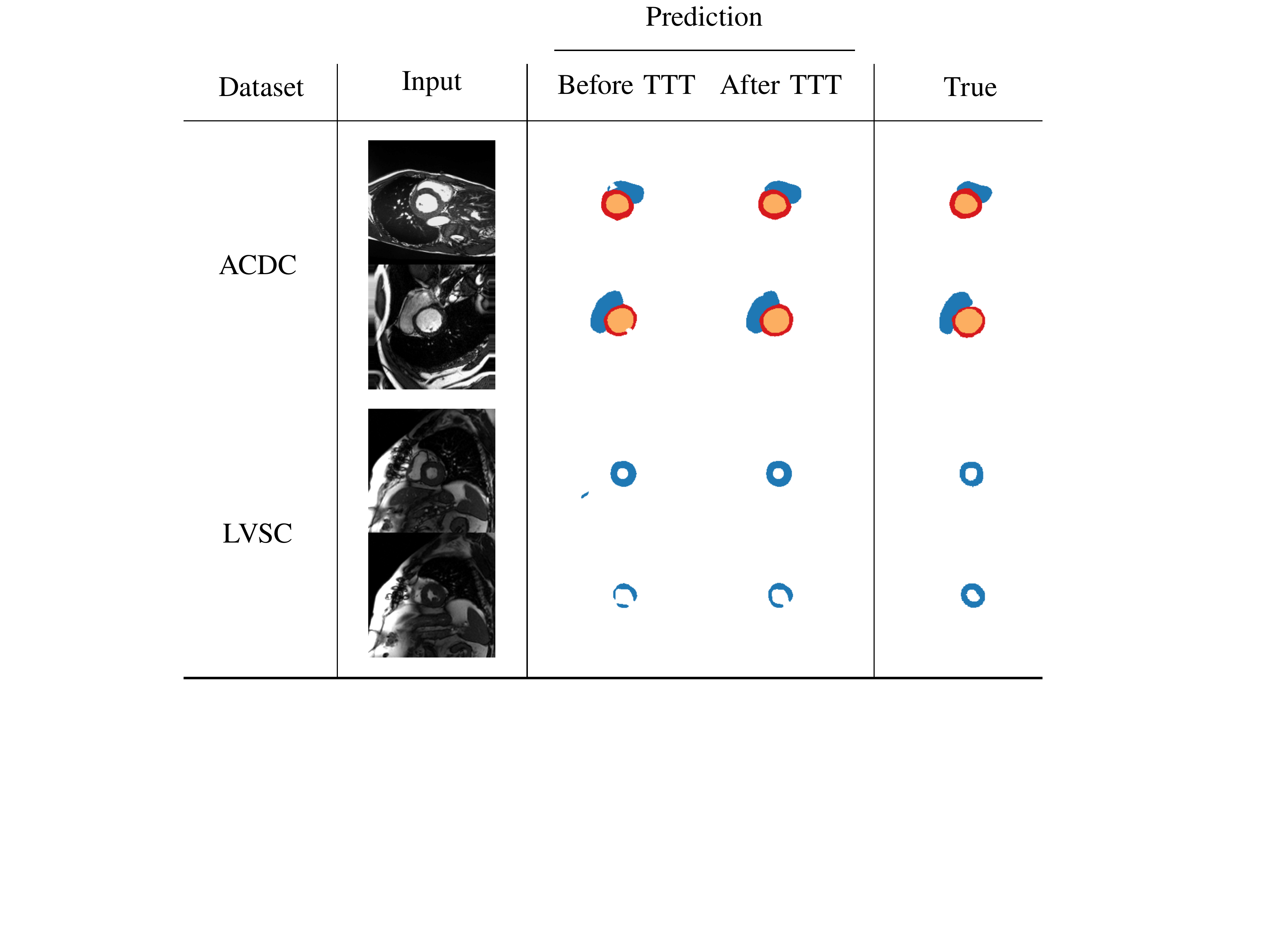}
    \caption{
    After adversarial Test-time Training, the segmentor corrects the initially erroneous segmentation masks to make them realistic, according to the adversarial shape prior.
    }
    \label{fig:acdc_test_samples}
\end{figure}

\clearpage

\section{Discriminator: Convergence and Memorisation}
We report examples of the training and validation losses for the GAN discriminator $\mathfunc{d}$. We use a Least-square GAN, whose discriminator loss to minimise is:
\begin{equation}\label{eq:discriminator_loss}
     \mathcal{V}_{LS}(\mathfunc{d}) = 
        \frac{1}{2} \underbrace{
            E_{\mathimage{y} \sim \mathdistrib{Y}}[(\mathfunc{d}(\mathimage{y}) - 1)^2]
        }_{\text{loss on \textit{real} samples}}
        + 
        \frac{1}{2} \underbrace{
            E_{\mathimage{x} \sim \mathdistrib{X}}[(\mathfunc{d}(\mathfunc{s}(\mathimage{x})) + 1)^2] 
        }_{\text{loss on \textit{fake} samples}}
\end{equation}
where $+1$ and $-1$ are the labels for \textit{real} and \textit{fake} (generated) images, respectively.

\begin{figure}[ht!]
    \centering
    \includegraphics[width=0.8\linewidth]{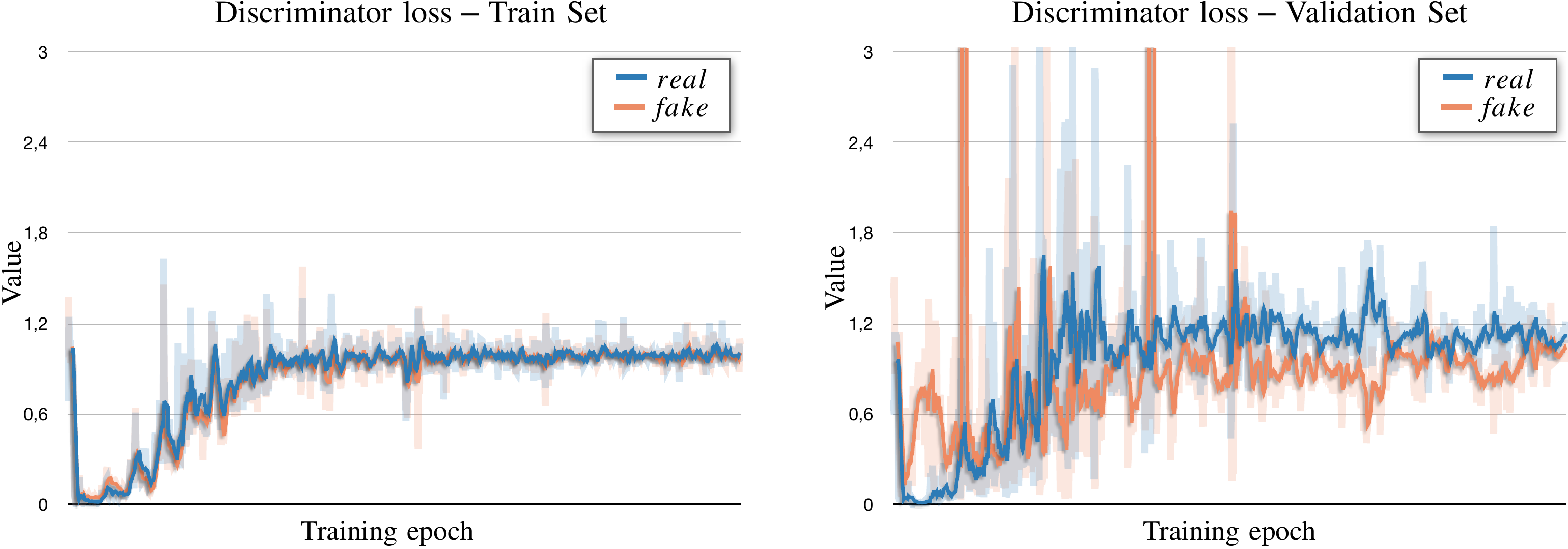}
    \caption{
    At convergence, the discriminator reaches an equilibrium stage where it always predicts the value 0, equidistant from the \textit{true} and the \textit{fake} labels. As a result, losses converge to the equilibrium value 1.0 both for train and validation.
    }
    \label{fig:gan_collapse_equilibrium}
\end{figure}

\begin{figure}[ht!]
    \centering
    \includegraphics[width=0.8\linewidth]{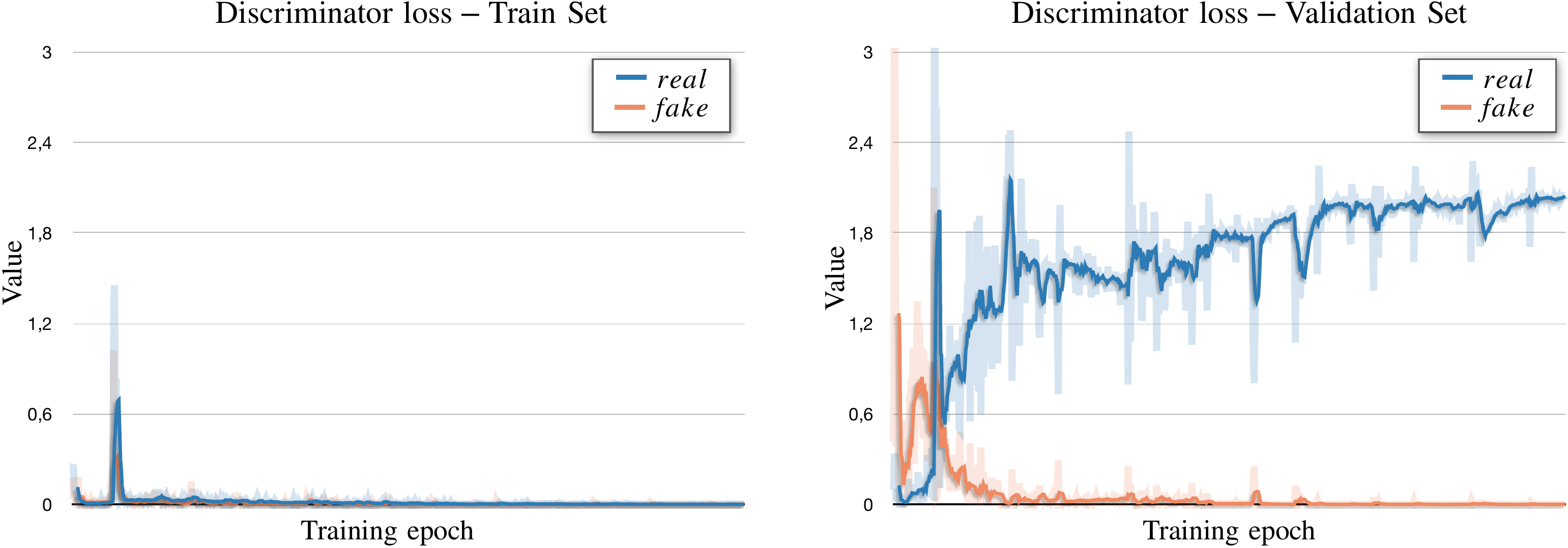}
    \caption{
    At convergence, the discriminator shows signals of memorisation. The discriminator memorises the \textit{real} training images, and it predicts the label \textit{fake} (i.e. the value -1) for any other case. During validation, the \textit{fake} images are still classified correctly, while the \textit{real} ones are classified as \textit{fake} and the associated loss converges to the value of 2.0. 
    }
    \label{fig:gan_collapse_memorisation}
\end{figure}

We report examples of convergence modes in Fig.~\ref{fig:gan_collapse_equilibrium} and Fig.~\ref{fig:gan_collapse_memorisation}. We show losses on the training set on the left, losses on the validation set on the right. Observe that $-$ despite the single loss components have different values $-$ the total loss $\mathcal{V}_{LS}(\mathfunc{d})$ on the validation set is the same in both cases.

